\documentclass[preprint,epsfig,onecolumn,floats, showkeys]{revtex4}
\usepackage{makeidx}
\usepackage{amsmath}
\usepackage{amssymb}
\usepackage{graphicx}
\usepackage{epsfig}

\begin{document}

\title{Loading N-Dimensional Vector into Quantum Registers from Classical
Memory with O(logN) Steps}
\thanks{The work is supported by the key item of science and technology
awarded by Science and Technology Bureau of Sichuan Province under Grant No.
2006J13-153, and the key item awarded by Sichuan Normal Univ. under Grant
No.06lk002}
\author{Pang Chao-Yang}
\email{cyp_900@hotmail.com}
\email{cypang@sicnu.edu.cn}
\affiliation{$^{1}$ Key Software Laboratory, Sichuan Normal University, Chengdu 610066,
China\\
$^{2}$ College of Mathematics and Software Science, Sichuan Normal
University, Chengdu 610066, China}

\begin{abstract}
Vector is the general format of input data of most algorithms. Designing
unitary operation to load all information of vector into quantum registers
of quantum CPU from classical memory is called quantum loading scheme (QLS).
QLS assembles classical memory and quantum CPU as a whole computer, which
will be important for further quantum computation. We present a QLS based on
path interference with time complexity $O(log_{2}N)$, while classical
loading scheme has time complexity $O(N)$, that is the efficiency bottleneck
of classical computer.
\end{abstract}

\maketitle

\section{Introduction}

Vector is the general format of input data of algorithm, and each component
of vector is stored in classical memory sequentially in general. Thus there
are two questions for general quantum algorithm, the first question is that
which state is suitable to represent all information of vector for further
quantum computation, and the second question is that how to load all
information of vector into quantum registers (or quantum state) without
losing information. Loading data set such as vector into classical registers
of CPU from classical memory is called \textbf{classical loading scheme (CLS)%
}. Similar to CLS, designing unitary operation to load all information of
vector into quantum registers of quantum CPU from classical memory is called
\textbf{quantum loading scheme (QLS)}. CLS or QLS assembles classical memory
and CPU as a whole computer. QLS\ makes quantum CPU is compatible with
classical memory.

An $N-$dimensional vector is denoted as $\vec{a}=\{a_{0},a_{1},...,a_{N-1}\}$%
, where the components $a_{0},a_{1},...,a_{N-1}$ are real numbers. It has
been shown that entangled state $\frac{1}{\sqrt{N}}({\sum\limits_{i=0}^{N-1}{%
{{\left\vert {i}\right\rangle }}}_{{register1}}{{{\left\vert a_{{i}%
}\right\rangle }}}_{{register2}}})${\ is suitable for the representation of
vector without losing any information of vector \cite{Pang
PostDocReport,Pang QVQ2,Pang QDCT,Pang QVQ1}. Let initial state ${\left\vert
{\phi _{0}}\right\rangle }$ be ${\left\vert {\phi _{0}}\right\rangle }={%
\left\vert {0}\right\rangle _{{{{q_{1}q_{2}...q_{n}}}}}\left\vert {0}%
\right\rangle }_{{_{{{{p_{1}p_{2}...p_{m}}}}}}}{\left\vert {ancilla_{1}}%
\right\rangle }$, where the $m+n$ qubits ${{{q_{1},...,q_{n},p_{1},...,p_{m}}%
}}$ are collected as a whole object, dividing them is prohibited, and where
the ancillary state ${\left\vert {ancilla_{1}}\right\rangle }$ is \emph{known%
}. QLS can be described as how to design unitary operation $%
U_{(0,1,...,N-1)} $ such that
\begin{equation}
{\left\vert {\phi }\right\rangle =U_{(0,1,...,N-1)}{\left\vert {\phi _{0}}%
\right\rangle }=}\frac{1}{\sqrt{N}}({\sum\limits_{i=0}^{N-1}{{{\left\vert {i}%
\right\rangle }}}_{{{{q_{1}q_{2}...q_{n}}}}}{{{\left\vert a_{{i}%
}\right\rangle }}}_{{{{p_{1}p_{2}...p_{m}}}}}}){\left\vert {ancilla_{2}}%
\right\rangle }\mathrm{,}  \label{eqTarget}
\end{equation}%
where $N=2^{n}$ and the ancillary state ${\left\vert {ancilla_{2}}%
\right\rangle }$ is \emph{known }}(all {ancillary states is \emph{known }}in
this paper.){. }

Nielsen and Chuang pointed out that quantum computer should have loading
scheme in principle to load classical database record into quantum registers
from classical database \cite[section 6.5]{Nielsen, QCZhao}. However, there
is no detailed work on QLS up till now. In fact, the research of QLS is
motivated by the quantum algorithm of image compression \cite{Pang
PostDocReport,Pang QVQ2,Pang QDCT,Pang QVQ1,Lattorre}. In this paper, we
present a QLS based on the path interference, which has been widely used in
quantum information processing, e.g. non-unitary computation\cite%
{Kwiat,LongGuiLu}. The unitary computation using path interference is
demonstrated in this paper, and the output of the unitary computation can be
measured with successful probability 100\% in theory. The time complexity of
our QLS is $O(log_{2}N)$, which exhibits a speed-up over CLS with time
complexity $O(N)$.

\section{The Design of QLS}

\subsection{Loading 2D Vector into Quantum Registers from Classical Memory}

\begin{figure}[h]
\epsfig{file=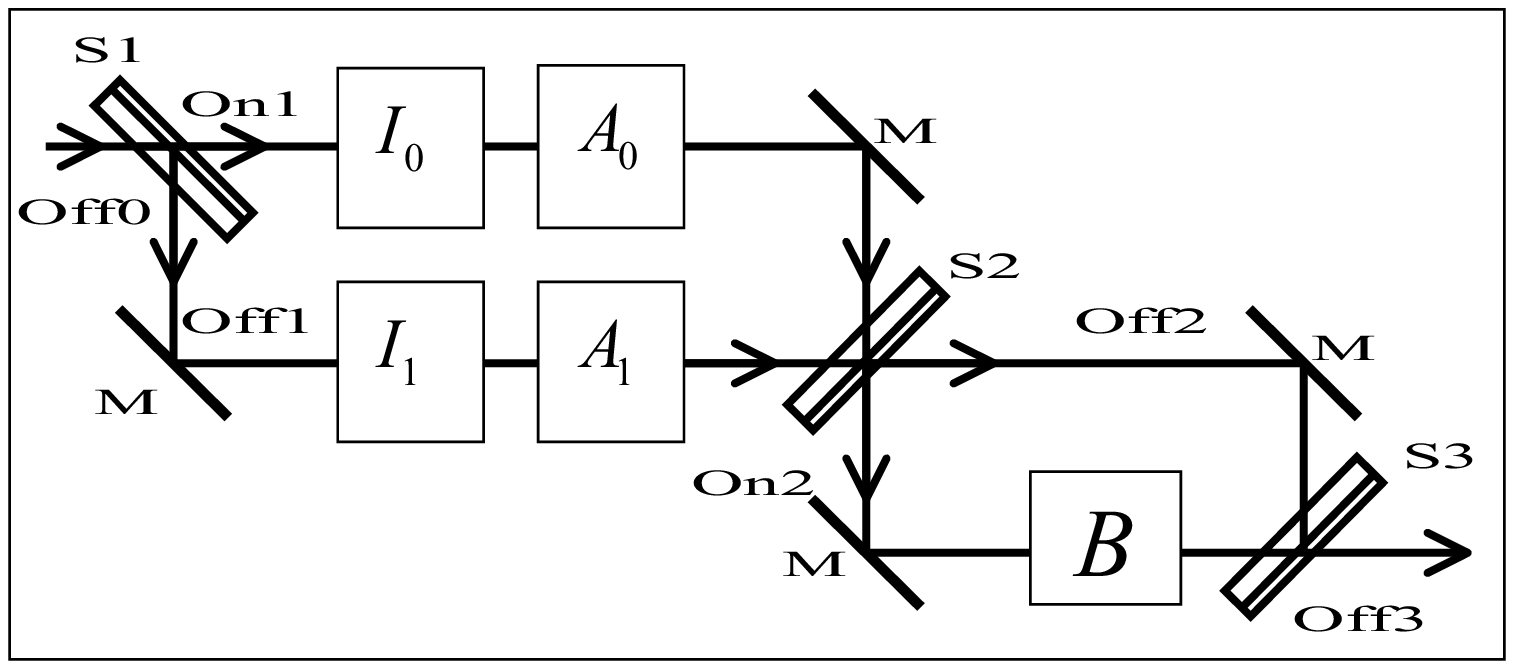,width=9cm,}
\caption{\textbf{The Illustration of the processing of unitary operation }$%
U_{(0,1)}$\textbf{\ that transforms state from }$\left\vert {0}\right\rangle
\left\vert {0}\right\rangle \left\vert {Off}_{{0}}\right\rangle $\textbf{\
into }$\frac{1}{\protect\sqrt{2}}(|0\rangle |a_{0}\rangle +|1\rangle
|a_{1}\rangle )\left\vert {Off}_{{3}}\right\rangle $. M: Mirror}
\label{figU01}
\end{figure}

The design of unitary operation $U_{(0,1)}$ that loads 2D vector $\vec{a}%
=\{a_{0},a_{1}\}$ is described conceptually as follows (see Fig.\ref{figU01}%
):

\textbf{Step 1} The switch $S_{1}$ applies rotation on the initial ancilla
state and transforms $\left\vert {Off_{0}}\right\rangle $ into%
\begin{equation*}
{\left\vert {Off_{0}}\right\rangle \overset{S_{1}}{\rightarrow }\frac{{{%
\left\vert {Off_{1}}\right\rangle }+{\left\vert {On_{1}}\right\rangle }}}{%
\sqrt{2}}}
\end{equation*}%
and generate the following state $\left\vert {\phi _{1}}\right\rangle $
\begin{equation}
{\left\vert {\phi _{1}}\right\rangle }{=}\frac{1}{\sqrt{2}}{\left\vert {0}%
\right\rangle }_{{{{q_{1}q_{2}...q_{n}}}}}{\left\vert {0}\right\rangle }_{{_{%
{{{p_{1}p_{2}...p_{m}}}}}}}{\left\vert {On_{1}}\right\rangle +}\frac{1}{%
\sqrt{2}}{\left\vert {0}\right\rangle }_{{{{q_{1}q_{2}...q_{n}}}}}{%
\left\vert {0}\right\rangle }_{{_{{{{p_{1}p_{2}...p_{m}}}}}}}{\left\vert {%
Off_{1}}\right\rangle }  \label{eqFai1}
\end{equation}

\textbf{Step 2 }Perform unitary operations $I_{0}$ and $A_{0}$ along `$%
On_{1} $' path, while perform unitary operations $I_{1}$ and $A_{1}$ along `$%
Off_{1} $' path.

\begin{equation*}
\begin{tabular}{lll}
$\left\{
\begin{tabular}{c}
${\left\vert {0}\right\rangle }_{{{{q_{1}q_{2}...q_{n}}}}}\overset{I_{0}}{%
\rightarrow }{\left\vert {0}\right\rangle }_{{{{q_{1}q_{2}...q_{n}}}}}$ \\
${{\left\vert {0}\right\rangle }_{{_{{{{p_{1}p_{2}...p_{m}}}}}}}}\overset{%
A_{0}}{{\rightarrow }}{{\left\vert a_{{0}}\right\rangle }_{{_{{{{%
p_{1}p_{2}...p_{m}}}}}}}}$%
\end{tabular}%
\right. $ & , & $\left\{
\begin{tabular}{c}
${\left\vert {0}\right\rangle }_{{{{q_{1}q_{2}...q_{n}}}}}\overset{I_{1}}{%
\rightarrow }{\left\vert {1}\right\rangle }_{{{{q_{1}q_{2}...q_{n}}}}}$ \\
${{\left\vert {0}\right\rangle }_{{_{{{{p_{1}p_{2}...p_{m}}}}}}}}\overset{%
A_{1}}{{\rightarrow }}{{\left\vert a_{{1}}\right\rangle }_{{_{{{{%
p_{1}p_{2}...p_{m}}}}}}}}$%
\end{tabular}%
\right. $%
\end{tabular}%
\end{equation*}

We assume the output of two pathes are simultaneous, then the state ${%
\left\vert {\phi _{2}}\right\rangle }$ is generated as

\begin{equation*}
\left\{
\begin{tabular}{c}
$\frac{1}{\sqrt{2}}{\left\vert {0}\right\rangle }_{{{{q_{1}q_{2}...q_{n}}}}}{%
\left\vert {0}\right\rangle }_{{_{{{{p_{1}p_{2}...p_{m}}}}}}}{\left\vert {%
On_{1}}\right\rangle }\overset{A_{0}I_{0}}{{\rightarrow }}\frac{1}{\sqrt{2}}{%
\left\vert {0}\right\rangle }_{{{{q_{1}q_{2}...q_{n}}}}}{\left\vert a_{{0}%
}\right\rangle }_{{_{{{{p_{1}p_{2}...p_{m}}}}}}}{\left\vert {On_{1}}%
\right\rangle }$ \\
$\frac{1}{\sqrt{2}}{\left\vert {0}\right\rangle }_{{{{q_{1}q_{2}...q_{n}}}}}{%
\left\vert {0}\right\rangle }_{{_{{{{p_{1}p_{2}...p_{m}}}}}}}{\left\vert {%
Off_{1}}\right\rangle }\overset{A_{1}I_{1}}{{\rightarrow }}\frac{1}{\sqrt{2}}%
{\left\vert {1}\right\rangle }_{{{{q_{1}q_{2}...q_{n}}}}}{\left\vert a_{{1}%
}\right\rangle }_{{_{{{{p_{1}p_{2}...p_{m}}}}}}}{\left\vert {Off_{1}}%
\right\rangle }$%
\end{tabular}%
\right.
\end{equation*}

\begin{equation}
\Rightarrow {\left\vert {\phi _{2}}\right\rangle =}\frac{1}{\sqrt{2}}{%
\left\vert {0}\right\rangle }_{{{{q_{1}q_{2}...q_{n}}}}}{\left\vert a_{{0}%
}\right\rangle }_{{_{{{{p_{1}p_{2}...p_{m}}}}}}}{\left\vert {On_{1}}%
\right\rangle +}\frac{1}{\sqrt{2}}{\left\vert {1}\right\rangle }_{{{{%
q_{1}q_{2}...q_{n}}}}}{\left\vert a_{{1}}\right\rangle }_{{_{{{{%
p_{1}p_{2}...p_{m}}}}}}}{\left\vert {Off_{1}}\right\rangle }  \label{eqFai2}
\end{equation}

The functions of $I_{0}$ and $I_{1}$ are to generate subscripts $0$ and $1$
respectively, and the functions of $A_{0}$ and $A_{1}$ are to generate
numbers $a_{0}$ and $a_{1}$ respectively. Because value $a_{0}$ and $a_{1}$
are both known numbers, flipping part of the $m+n$ qubits ${{{%
q_{1},...,q_{n},p_{1},...,p_{m}}}}$ will generate states ${{\left\vert a_{{0}%
}\right\rangle }_{{_{{{{p_{1}p_{2}...p_{m}}}}}}}}$ or ${{\left\vert a_{{1}%
}\right\rangle }_{{_{{{{p_{1}p_{2}...p_{m}}}}}}}}$. Thus, the unitary
operations $I_{0}$, $I_{1}$, $A_{0}$, $A_{1}$ is easy to be designed.

\textbf{Step 3 }The switch $S_{2}$ applies rotation on the initial ancilla
state {as}

\begin{equation*}
\left\{
\begin{tabular}{c}
${\left\vert {Off_{1}}\right\rangle }\overset{S_{2}}{\rightarrow }{\frac{{{%
\left\vert {Off_{2}}\right\rangle }-{\left\vert {On_{2}}\right\rangle }}}{%
\sqrt{2}}}$ \\
${\left\vert {On_{1}}\right\rangle }\overset{S_{2}}{\rightarrow }{\frac{{{%
\left\vert {Off_{2}}\right\rangle }}+{{\left\vert {On_{2}}\right\rangle }}}{%
\sqrt{2}}}$%
\end{tabular}%
\right.
\end{equation*}%
and generate the following state ${\left\vert {\phi _{3}}\right\rangle }$

\begin{eqnarray}
{\left\vert {\phi _{3}}\right\rangle } &=&\frac{1}{2}({\left\vert {0}%
\right\rangle }_{{{{q_{1}q_{2}...q_{n}}}}}{\left\vert a_{{0}}\right\rangle }%
_{{_{{{{p_{1}p_{2}...p_{m}}}}}}}+{\left\vert {1}\right\rangle }_{{{{%
q_{1}q_{2}...q_{n}}}}}{\left\vert a_{{1}}\right\rangle }_{{_{{{{%
p_{1}p_{2}...p_{m}}}}}}}){\left\vert {Off_{2}}\right\rangle }  \notag \\
&&{+}\frac{1}{2}({\left\vert {0}\right\rangle }_{{{{q_{1}q_{2}...q_{n}}}}}{%
\left\vert a_{{0}}\right\rangle }_{{_{{{{p_{1}p_{2}...p_{m}}}}}}}-{%
\left\vert {1}\right\rangle }_{{{{q_{1}q_{2}...q_{n}}}}}{\left\vert a_{{1}%
}\right\rangle }_{{_{{{{p_{1}p_{2}...p_{m}}}}}}}){\left\vert {On_{2}}%
\right\rangle }  \label{eqFai3}
\end{eqnarray}

\textbf{Step 4 }Apply phase transformaition $B$ along `$On_{2}$' path.

\begin{equation}
B={\left\vert {0}\right\rangle \left\vert a_{{0}}\right\rangle }\langle a_{{0%
}}|\langle 0|-{\left\vert {1}\right\rangle \left\vert a_{{1}}\right\rangle }%
\langle a_{{1}}|\langle 1|  \label{eqB}
\end{equation}

It's a very fast operation and generates the state ${\left\vert {\phi _{4}}%
\right\rangle }$

\begin{equation}
{\left\vert {\phi _{4}}\right\rangle }=\frac{1}{\sqrt{2}}({\left\vert {0}%
\right\rangle }_{{{{q_{1}q_{2}...q_{n}}}}}{\left\vert a_{{0}}\right\rangle }%
_{{_{{{{p_{1}p_{2}...p_{m}}}}}}}+{\left\vert {1}\right\rangle }_{{{{%
q_{1}q_{2}...q_{n}}}}}{\left\vert a_{{1}}\right\rangle }_{{_{{{{%
p_{1}p_{2}...p_{m}}}}}}})(\frac{{{\left\vert {Off_{2}}\right\rangle }}+{{%
\left\vert {On_{2}}\right\rangle }}}{\sqrt{2}})  \label{eqFai4}
\end{equation}

\textbf{Step 5 }The switch $S_{3}$ applies rotation on the initial ancilla
state {as}

\begin{equation*}
\left\{
\begin{tabular}{c}
${\left\vert {Off_{2}}\right\rangle }\overset{S_{3}}{\rightarrow }{\frac{{{%
\left\vert {Off_{3}}\right\rangle }+{\left\vert {On_{3}}\right\rangle }}}{%
\sqrt{2}}}$ \\
${\left\vert {On_{2}}\right\rangle }\overset{S_{3}}{\rightarrow }{\frac{{{%
\left\vert {Off_{3}}\right\rangle }}-{{\left\vert {On_{3}}\right\rangle }}}{%
\sqrt{2}}}$%
\end{tabular}%
\right.
\end{equation*}%
and generate the final state ${\left\vert {\phi }\right\rangle }$

\begin{equation}
{\left\vert {\phi }\right\rangle }=\frac{1}{\sqrt{2}}({\left\vert {0}%
\right\rangle }_{{{{q_{1}q_{2}...q_{n}}}}}{\left\vert a_{{0}}\right\rangle }%
_{{_{{{{p_{1}p_{2}...p_{m}}}}}}}+{\left\vert {1}\right\rangle }_{{{{%
q_{1}q_{2}...q_{n}}}}}{\left\vert a_{{1}}\right\rangle }_{{_{{{{%
p_{1}p_{2}...p_{m}}}}}}}){{\left\vert {Off_{3}}\right\rangle }}
\label{eqFai}
\end{equation}

Fig.\ref{figU01} and Eq.(\ref{eqU01}) illustrate the processing of operation
$U_{(0,1)}$.

\begin{equation}
\begin{tabular}{c}
$|0\rangle |0\rangle |Off_{0}\rangle \overset{S_{1}}{\rightarrow }%
\left\langle
\begin{tabular}{c}
$\frac{1}{\sqrt{2}}|0\rangle |0\rangle |On_{1}\rangle \overset{A_{0}I_{0}}{%
\rightarrow }\frac{1}{\sqrt{2}}|0\rangle |a_{0}\rangle |On_{1}\rangle $ \\
$\frac{1}{\sqrt{2}}|0\rangle |0\rangle |Off_{1}\rangle \overset{A_{1}I_{1}}{%
\rightarrow }\frac{1}{\sqrt{2}}|1\rangle |a_{1}\rangle |Off_{1}\rangle $%
\end{tabular}%
\right\rangle \overset{S_{2}}{\rightarrow }$ \\
$\left\langle
\begin{tabular}{c}
$\frac{1}{2}(|0\rangle |a_{0}\rangle +|1\rangle |a_{1}\rangle
)|Off_{2}\rangle $ \\
$\frac{1}{2}(|0\rangle |a_{0}\rangle -|1\rangle |a_{1}\rangle )|On2\rangle
\overset{B}{\rightarrow }\frac{1}{2}(|0\rangle |a_{0}\rangle +|1\rangle
|a_{1}\rangle )|On_{2}\rangle $%
\end{tabular}%
\right\rangle \overset{S_{3}}{\rightarrow }{\left\vert {\phi }\right\rangle }
$%
\end{tabular}
\label{eqU01}
\end{equation}%
Here as well as in the following discussions, all subscripts of registers
are ignored.

\subsection{Loading 4D Vector or Multi-Dimensional into Quantum Registers}

\begin{figure}[h]
\epsfig{file=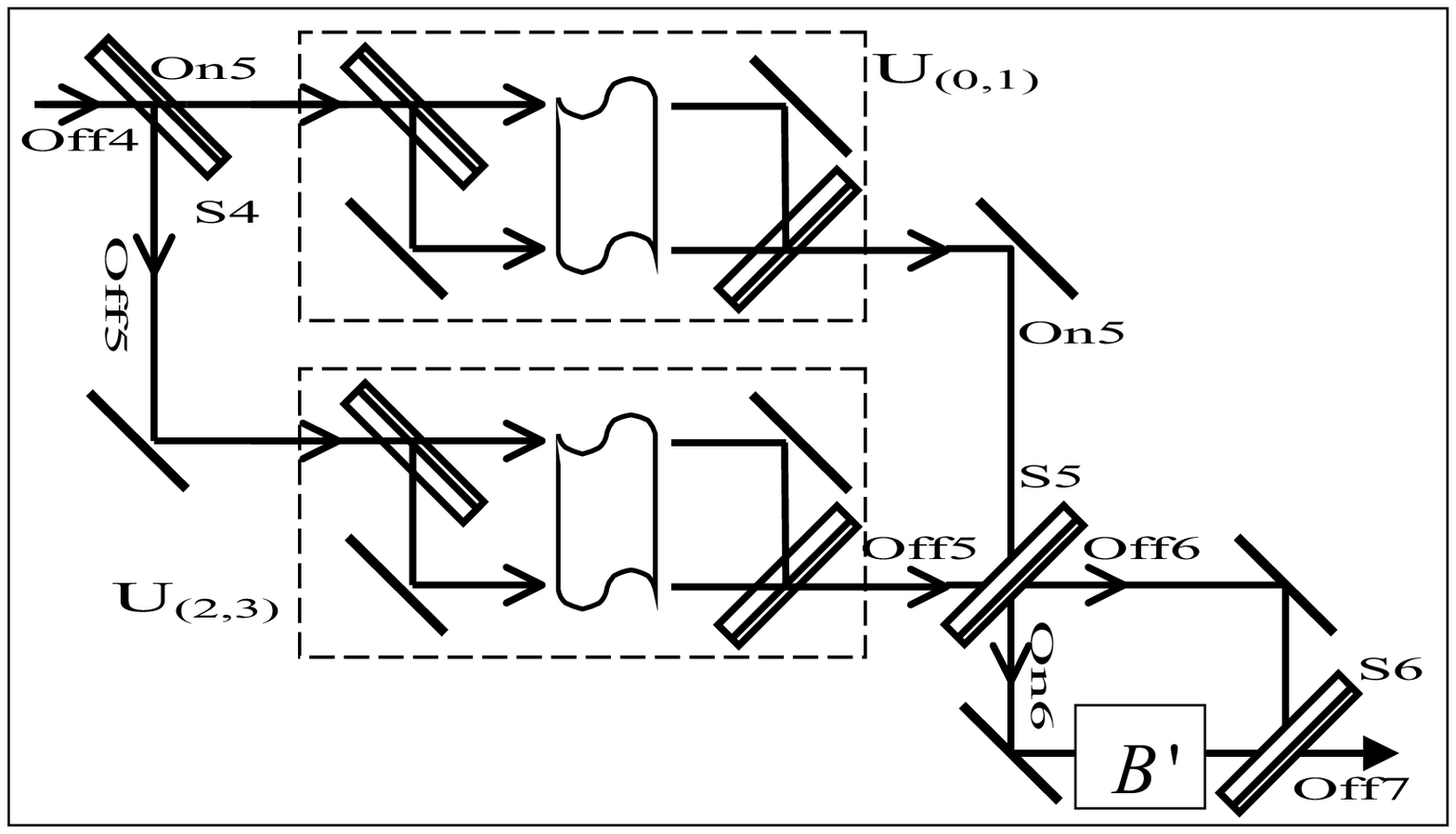,width=8cm,}
\caption{\textbf{The illustration of unitary operation }$U_{(0,1,2,3)}$%
\textbf{\ that transforms state from }$\left\vert {0}\right\rangle
\left\vert {0}\right\rangle \left\vert {Off}_{{4}}\right\rangle $\textbf{\
into }$(\protect\underset{i=0}{\protect\overset{3}{\sum }}\frac{1}{2}%
\left\vert {i}\right\rangle \left\vert a_{{i}}\right\rangle )\left\vert {Off}%
_{{7}}\right\rangle $.}
\label{figU0123}
\end{figure}

The design of unitary operation $U_{(0,1,2,3)}$ is described conceptually as
follows (see Fig.\ref{figU0123}):

\begin{description}
\item[Step 1] Construct unitary operation $S_{4}$, $S_{5}$, $S_{6},B^{\prime
}$ as following:
\end{description}

$\left\{
\begin{tabular}{c}
${\left\vert {Off}_{i}\right\rangle }\overset{S_{i}}{\rightarrow }{\frac{{{%
\left\vert {Off_{i+1}}\right\rangle }+{\left\vert {On_{i+1}}\right\rangle }}%
}{\sqrt{2}}}$ \\
${\left\vert {On}_{i}\right\rangle }\overset{S_{i}}{\rightarrow }{\frac{{{%
\left\vert {Off_{i+1}}\right\rangle }}-{{\left\vert {On_{i+1}}\right\rangle }%
}}{\sqrt{2}}}$%
\end{tabular}%
\right. $, $\left\{
\begin{tabular}{c}
${\left\vert {Off}_{5}\right\rangle }\overset{S_{5}}{\rightarrow }{\frac{{{%
\left\vert {Off_{i+1}}\right\rangle }-{\left\vert {On_{i+1}}\right\rangle }}%
}{\sqrt{2}}}$ \\
${\left\vert {On}_{5}\right\rangle }\overset{S_{5}}{\rightarrow }{\frac{{{%
\left\vert {Off_{i+1}}\right\rangle }}+{{\left\vert {On_{i+1}}\right\rangle }%
}}{\sqrt{2}}}$%
\end{tabular}%
\right. $, $\ B^{\prime }=|\alpha \rangle \langle \alpha |-|\beta \rangle
\langle \beta |$,

where $i=4,6$, $|\alpha \rangle =\frac{1}{\sqrt{2}}(|0\rangle |a_{0}\rangle
+|1\rangle |a_{1}\rangle )$ and $|\beta \rangle =\frac{1}{\sqrt{2}}%
(|2\rangle |a_{2}\rangle +|3\rangle |a_{3}\rangle )$

\begin{description}
\item[Step 2] Assemble unitary operations $S_{4}$, $S_{5}$, $S_{6},B^{\prime
},U_{(0,1)}$ and $U_{(2,3)}$ according to Fig.\ref{figU0123} to form unitary
operations $U_{(0,1,2,3)}$.
\end{description}

Eq.(\ref{eqU0123}) illustrates the processing of operation $U_{(0,1,2,3)}$.

\begin{equation}
\begin{tabular}{c}
$|0\rangle |0\rangle |Off_{4}\rangle \overset{S_{4}}{\rightarrow }%
\left\langle
\begin{tabular}{c}
$\frac{1}{\sqrt{2}}|0\rangle |0\rangle |On_{5}\rangle \overset{U_{(0,1)}}{%
\rightarrow }\frac{1}{2}(|0\rangle |a_{0}\rangle +|1\rangle |a_{1}\rangle
)|On_{5}\rangle $ \\
$\frac{1}{\sqrt{2}}|0\rangle |0\rangle |Off_{5}\rangle \overset{U_{(2,3)}}{%
\rightarrow }\frac{1}{2}(|2\rangle |a_{2}\rangle +|3\rangle |a_{3}\rangle
)|Off_{5}\rangle $%
\end{tabular}%
\right\rangle \overset{S_{5}}{\rightarrow }$ \\
$\left\langle
\begin{tabular}{c}
$\frac{1}{2}[\frac{1}{\sqrt{2}}(|0\rangle |a_{0}\rangle +|1\rangle
|a_{1}\rangle )+\frac{1}{\sqrt{2}}(|2\rangle |a_{2}\rangle +|3\rangle
|a_{3}\rangle )]|Off_{6}\rangle $ \\
\\
$\frac{1}{2}[\frac{1}{\sqrt{2}}(|0\rangle |a_{0}\rangle +|1\rangle
|a_{1}\rangle )-\frac{1}{\sqrt{2}}(|2\rangle |a_{2}\rangle +|3\rangle
|a_{3}\rangle )]|On_{6}\rangle \overset{B^{\prime }}{\rightarrow }$ \\
$\frac{1}{2}[\frac{1}{\sqrt{2}}(|0\rangle |a_{0}\rangle +|1\rangle
|a_{1}\rangle )+\frac{1}{\sqrt{2}}(|2\rangle |a_{2}\rangle +|3\rangle
|a_{3}\rangle )]|On_{6}\rangle $%
\end{tabular}%
\right\rangle \overset{S_{6}}{\rightarrow }{\left\vert {\phi }\right\rangle }
$%
\end{tabular}
\label{eqU0123}
\end{equation}

If the unitary operations $U_{(0,1)}$ and $U_{(2,3)}$ embedded in Fig.\ref%
{figU0123} are replaced by $U_{(0,1,2,3)}$ and $U_{(4,5,6,7)}$ respectively,
then $U_{(0,1,...,7)}$ is constructed. Similar to Fig.\ref{figU0123}, we can
apply the same method to construct unitary operation $U_{(0,1,...,2^{n})}$.
If $N\neq 2^{n}$, we could add extra zero components to create a $2^{n}$%
-dimensional vector.

\subsection{Loading Vector into State $\frac{1}{\protect\sqrt{N}}({%
\sum\limits_{i=0}^{N-1}{{{\left\vert {i}\right\rangle \left\vert
0\right\rangle )}}}}$ to Form Entangled State $\frac{1}{\protect\sqrt{N}}({%
\sum\limits_{i=0}^{N-1}{{{\left\vert {i}\right\rangle \left\vert a_{{i}%
}\right\rangle )}}}}$}

Grover's algorithm \cite{Nielsen} has the function that find the index $%
i_{0} $ of a special database record $record_{i_{0}}$ from the index
superposition of state $\frac{1}{\sqrt{N}}({\sum\limits_{i=0}^{N-1}{{{%
\left\vert {i}\right\rangle )}}}}$ taking $O(\sqrt{N})$ steps. And the
record $record_{i_{0}}$\ is the genuine answer wanted by us. However, the
corresponding record $record_{i_{0}}$ can not be measured out unless the
1-1mapping relationship between index $i$ and the corresponding record $%
record_{i}$ is bound in the entangled state $\frac{1}{\sqrt{N}}({%
\sum\limits_{i=0}^{N-1}{{{\left\vert {i}\right\rangle {{{\left\vert
record_{i}\right\rangle }}})}}}}$. That is, we need a unitary$\ $operation $%
U_{L}$ such that
\begin{equation}
\frac{1}{\sqrt{N}}({\sum\limits_{i=0}^{N-1}{{{\left\vert {i}\right\rangle
\left\vert 0\right\rangle )\left\vert {ancilla_{4}}\right\rangle }}}}\overset%
{U_{L}}{{\rightarrow }}\frac{1}{\sqrt{N}}({\sum\limits_{i=0}^{N-1}{{{%
\left\vert {i}\right\rangle \left\vert a_{{i}}\right\rangle )\left\vert {%
ancilla_{3}}\right\rangle }}}}  \label{eqUL}
\end{equation}

Ref.\cite{Pang QVQ1, Pang QDCT} generalize Grover's algorithm to the general
search case with complex computation, and $U_{L}$ is required in this
general search case.

\begin{figure}[h]
\caption{\textbf{The Illustration of Unitary Operation }$(U_{L})^{\dagger }$%
: $\frac{1}{\protect\sqrt{2}}({\sum\limits_{i=0}^{1}{{{\left\vert {i}%
\right\rangle \left\vert a_{{i}}\right\rangle )\left\vert {Off_{3}}%
\right\rangle }}}}\rightarrow \frac{1}{\protect\sqrt{2}}({%
\sum\limits_{i=0}^{1}{{{\left\vert {i}\right\rangle \left\vert
0\right\rangle )\left\vert {Off}\right\rangle }}}}$. Operation $U_{L}$ can
be designed using the same method shown in Fig.\protect\ref{figU01} and Fig.%
\protect\ref{figU0123}. $S_{0}$: ${\left\vert {Off_{0}}\right\rangle }%
\rightarrow \frac{1}{\protect\sqrt{2}}({{\left\vert {Off}\right\rangle }+{%
\left\vert {On}\right\rangle }})${, }${\left\vert {On_{0}}\right\rangle }%
\rightarrow \frac{1}{\protect\sqrt{2}}({{\left\vert {Off}\right\rangle }}-{{%
\left\vert {On}\right\rangle }})$. Phase transformation $D={\left\vert i_{{1}%
}\right\rangle \left\vert 0\right\rangle }\langle 0|\langle i_{{1}}|-{%
\left\vert i_{{0}}\right\rangle \left\vert 0\right\rangle }\langle 0|\langle
i_{{0}}|$, where $i_{{0}}=0$, $i_{{1}}=1$.}
\label{figUL}\epsfig{file=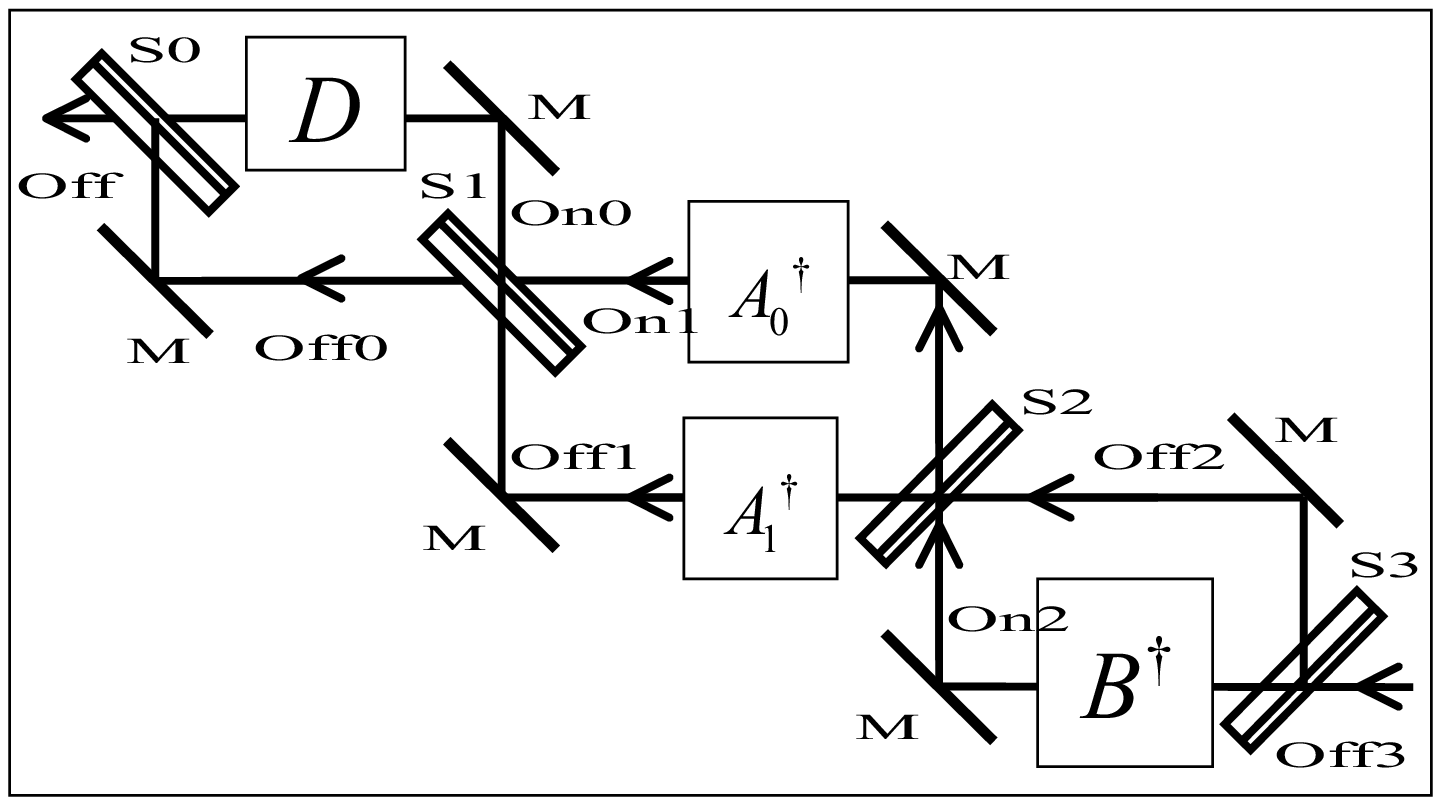,width=8cm,}
\end{figure}

$U_{L}$ can be designed using the same method shown in Fig.\ref{figU01} and
Fig.\ref{figU0123}. Fig.\ref{figUL} shows the design of the inverse unitary
operation $(U_{L})^{\dagger }$ at the case $N=2$. $U_{L}$ has time
complexity $O(log_{2}N)$ (unit time: phase transformation and flipping the
qubits of registers).

It has been demonstrated that giant molecules, such as charcoal $c_{60}$,
exhibit quantum interference \cite{Zeilinger}. Thus many freedom degrees of
giant molecule can be regarded as qubits to realize the QLS presented in
this paper. In addition, one of QLS application is that QLS can load the
data of image with huge size into quantum registers at a time for further
image compression \cite{Pang PostDocReport,Pang QVQ2,Pang QDCT,Pang QVQ1},
while only one data can be loaded into registers at a time for classical
computer.

\section{Conclusion}

Designing simple and fast unitary operation to load classical data set, such
as vector, into quantum registers from classical memory is called quantum
loading scheme (QLS). QLS\ makes quantum CPU is compatible with classical
memory, and it assembles classical memory and quantum CPU as a whole. QLS is
the base of further quantum computation. The QLS with time complexity $%
O(log_{2}N)$ (unit time: phase transformation and flipping the qubits of
registers)\ is presented in this paper, while classical loading scheme (CLS)
has time complexity $O(N)$ (unit: addition) because all computation
instructions have to be executed one by one. Path interference is applied to
design QLS in this paper so that the complexity of designing quantum
algorithm is decomposed as the design of many simple unitary operations. In
addition, this paper demonstrates that using path interference to design
unitary operation and parallel quantum computation is possible.

\begin{acknowledgments}
The author thanks Dr. Z.-W Zhou\ who is at Key Lab. of Quantum Information,
Uni. of Science and Technology of China for that he points out two errors in
author's primary idea. The author's first error is that the result generate
with probability 50\% for 2D vector, the second error is the defect of Fig.%
\ref{figU0123} that the output is direct product state. Dr. Z.-W Zhou tries
his best to help author for nearly 3 years. The author thanks his teacher,
prof. G.-C Guo. The author is brought up from Guo's Lab.. The author thanks
prof. V. N. Gorbachev who is at St.-Petersburg State Uni. of Aerospace
Instrumentation for the useful discussion with him. The author thanks Mir.
N. Kiesel who is at Max-Plank-Institute fur Quantenoptik, Germany for his
checking the partial deduction of section 2 of this paper. The author thanks
prof. G.-L Long who is at Tsinghua Uni., China for the useful discussion
with him and the author obtains some heuristic help from his eprint file
quant-ph/0512120. The author thanks associate prof. Shiuan-Huei Lin who is
at National Chiao Tung Uni., Taiwan., China for encouraging the author. The
author thanks prof. Hideaki Matsueda who is at Kochi Uni., Japan for
encouraging the author. The author thanks prof. J. Zhang and B.-P Hou who
are at Sichuan Normal Uni., China for their help. The author thanks prof.
Z.-F. Han, Dr. Y.-S. Zhang, Dr. Y.-F. Huang, Dr. Y.-J. Han, Mr. J.-M Cai,
Mr. M.-Y. Ye, and Mr. M.-Gong for their help and suggestions. One of
reviewers presents many significative suggestions to improve the readability
of this paper, the author thanks the reviewer.
\end{acknowledgments}


\begin{thebibliography}{99}
\bibitem{Pang PostDocReport} Pang C Y 2006 Quantum Image Compression
(Postdoctoral Report) (Heifei: University of Science and Technology of China)

\bibitem{Pang QVQ2} Pang C Y, Zhou Z W, and Guo G C 2006 \textit{Chin. Phys.}
\textbf{15} 3039

\bibitem{Pang QDCT} Pang C Y, Zhou Z W, and Guo G C 2006\
arXiv:quant-ph/0601043

\bibitem{Pang QVQ1} Pang C Y, Zhou Z W, Chen P X, and Guo G C 2006 \textit{%
Chin. Phys.} \textbf{15} 618

\bibitem{Lattorre} Latorre J I 2005 arXiv:quant-ph/0510031

\bibitem{Nielsen} Nielsen M A and Chuang I L 2000 Quantum Computationand and
Quantum Information (London: Cambridge University Press)

\bibitem{QCZhao} Zhao Q C(translator), Nielsen M A and Chuang I L 2004
Quantum Computation and Quantum Information (Beijing: Tsinghua University
publishers) (in Chinese)

\bibitem{Kwiat} Hosten O, Rakher M T, Barreiro J T, Peters N A, Kwiat P G
2006 \textit{Nature} \textbf{439} 949

\bibitem{LongGuiLu} Long G\ L 2006 \textit{Communications in Theoretical
Physics} \textbf{45} 825 .

\bibitem{Zeilinger} Arndt M, Nairz O, Vos-Andreae J, Keller C, van der Zouw
G, and Zeilinger A. 1999 \textit{Nature}(London), \textbf{401} 680
\end{thebibliography}
\end{document}